\documentclass[iop,numberedappendix]{emulateapj}
\usepackage{amsmath,amssymb,graphicx,xspace,color}
\usepackage{bm}
\usepackage{times}

\newcommand{\Fig}[1]{Figure~\ref{#1}}
\newcommand{\Eq}[1]{Equation~(\ref{#1})}
\newcommand{\EQ}{\begin{equation}}
\newcommand{\EN}{\end{equation}}
\newcommand{\vv}{\mbox{\boldmath $v$} {}}
\newcommand{\uu}{\mbox{\boldmath $u$} {}}
\newcommand{\FF}{\mbox{\boldmath $F$} {}}
\newcommand{\DD}{\mbox{\boldmath $D$} {}}

\newcommand{\Tab}[1]{Table~\ref{#1}}
\newcommand{\St}{\text{St}}

\newcommand{\rhoS}{\rho_{{\bullet}}}

\newcommand{\CFS}{\text{CF}_S}
\newcommand{\CFP}{\text{CF}_p}
\newcommand{\CFT}{\text{CF}_T}
\newcommand{\bS}{\bar{S}}
\newcommand{\vth}{v_{\text{th}}}
\newcommand{\vthM}{v_{\text{th}M}}
\newcommand{\vthO}{v_{\text{th}0}}

\graphicspath{{.}{./Figures/}}

\date{\today,~ $ $Revision: 1.20 $ $}
\begin{document}

\title{Dust concentration and chondrule formation}
\shorttitle{Dust concentration and chondrule formation}

\author{Alexander Hubbard\altaffilmark{1}, Mordecai-Mark Mac Low\altaffilmark{1}, Denton S. Ebel\altaffilmark{2}}
\altaffiltext{1}{Dept.~of Astrophysics, American Museum of Natural History, New York, NY, USA}
\altaffiltext{2}{Dept.~of Earth and Planetary Sciences, American Museum of Natural History, New York, NY, USA}
\email{{\tt ahubbard@amnh.org}}

\begin{abstract}
Meteoritical and astrophysical models of planet formation make contradictory predictions for dust concentration factors in chondrule forming
regions of the solar nebula.
Meteoritical and cosmochemical models strongly suggest that chondrules, a key component of the meteoritical record, formed in regions
with solids-to-gas mass ratios orders of magnitude above background. However, models of dust grain dynamics in protoplanetary disks struggle to surpass factors
of a few outside of very brief windows in the lifetime of the dust grains. Worse, those models do not predict significant concentration factors for
dust grains the size of chondrule precursors. We briefly develop the difficulty in concentrating dust particles
in the context of nebular chondrule formation and show that the disagreement is sufficiently
stark that cosmochemists should explore ideas that might revise the concentration factor requirements downwards.
\end{abstract}



\section{Introduction}

Chondrules are sub-mm melted glassy beads found within chondritic meteorites, generally, although not universally \citep{1953GeCoA...4...36U,
2014ApJ...794...91D}, thought
to have been generated by melting pre-floating dust grains in the solar nebula.
Our understanding of chondritic meteorites and the chondrules they contain remains spectacularly incomplete. Just one of the many fascinating challenges
they pose to our understanding of the formation of our solar system and extrasolar planetary systems is the question of dust concentration.
Cosmochemists studying chondrules find that they had to have melted in high density regions immensely (factors of hundreds, \citealt{1963Icar....2..152W,2006mess.book..253E}) enriched
in condensables above the expected solar ratio \citep{2000GeCoA..64..339E,2015GeCoA.148..228T}. Stabilizing liquids at chondrule melting temperatures
requires high partial pressures \citep{1967GeCoA..31.2095W}.
Planetesimal formation and dust coagulation theorists on the other hand wrestle with the need
for condensable enrichments of factors of a few
\citep{1995Icar..114..237D, 2016SSRv..205...41B}, and find that when such occur, they require
grains larger than chondrules or their precursors, and are generally short lived \citep{2016arXiv161107014Y}.

The question of particle concentration faces two complementary
difficulties. On one hand, at the densities assumed in cosmochemical calculations,
comparable to or higher than the Minimum Mass Solar Nebula \citep{1981PThPS..70...35H},
chondrules and their precursors are small and difficult to concentrate. On the other hand, as the disk evolves 
gas loss in accretion or winds means that the disk density drops.
When the dissipated gas is sufficiently rarefied to allow the concentration of objects the size of chondrule and chondrule precursors, those concentrations
are expected to rapidly lead to planetesimal formation \citep{2007Natur.448.1022J},
and thus be a brief stage in the life time of the dust. In that scenario, one must explain
how chondrule formation could strike those narrow windows so precisely.

We adapt dust transport theory for the purposes of the concentration of chondrules and their precursors, and
show how difficult it is to achieve the requested dust concentration factors.
This argues for new avenues of research on ways to reduce the required concentration factors.
The isotopic measurements of \cite{2016PNAS..113.2886B} demonstrate that chondrule melting
regions likely contained different material than chondrite assembly regions, and \cite{2016GeCoA.172..322E} show that
chondrules have distinctly different overall compositions than the rest of their host meteorite, in particular, less iron.
\cite{2011P&SS...59.1888E} found major differences in the chondrule chemical outcomes between 
cases of melting in regions enriched in CI (bulk chondritic meteoritic) material
versus melting in regions enriched in chondritic-IDP-like material although the concentration factors considered were on the order of thousands.
Similarly, chondrule formation events need not have been in equilibrium as assumed in the calculations of, e.g.~\cite{2000GeCoA..64..339E}.
The composition of chondrules might have been kinetically controlled \citep{1996GeCoA..60.1445N}
as long as cooling rates were sufficiently high to cool newly formed chondrules within tens of minutes to hours \citep{2002M&PS...37..183D}.

\section{Mechanics of dust concentration}
\label{MDC}

Protoplanetary disks are expected to support some level of turbulence. Inertial particles are not strictly passive tracers,
and turbulent, rather than microphysical, diffusion does have some non-diffusive consequences. For example, preferential concentration concentrates inertial
dust grains in high shear-regions between turbulent vortices \citep{1987JFM...174..441M,
2001ApJ...546..496C}, while  turbulent thermal diffusion, perhaps ill-named,
pumps inertial grains from hot regions to cold ones \citep{2016MNRAS.456.3079H}.
However, preferential concentration relies on aerodynamically identical particles, and protoplanetary disks have
too weak global thermal gradients to be good
hosts for  turbulent thermal diffusion, so we can approximate dust concentration as the balance between pressure maxima concentrating dust and turbulent diffusion
smoothing out those dust concentrations. 

Small dust grains in protoplanetary disks are frictionally entrained by the gas with a drag acceleration
\EQ
\left.\frac{\partial \vv}{\partial t}\right|_\text{drag} = -\frac{\vv-\uu}{\tau}  \label{dvdt}
\EN
where $\vv$ is the velocity of a dust grain, $\uu$ the velocity of the gas at the dust grain's position, and $\tau$ the dust
grain's frictional stopping time.
Grains with radii smaller than the local gas mean free path (which includes chondrules and, absent extreme porosity, their precursors)
are in the Epstein drag regime with
\EQ
\tau = \frac{a \rhoS}{\rho_g \vth}, \label{tauE}
\EN
where $a$ and $\rhoS$ are the dust grain's radius and solid density, while $\rho_g$ is the gas density.
The gas thermal speed $\vth$ is given by
\EQ
\vth \equiv \sqrt{\frac{8 k_B T}{\pi m_g}} \label{vth},
\EN
where $m_g \simeq 2.3$\,amu is the gas mean molecular weight.
We also define the Stokes number
\EQ
\St \equiv \tau \Omega, \label{St_def}
\EN
where $\Omega$ is the local orbital frequency. Dust grains with $\St \ll 1$, which include chondrules and their
precursors, are well coupled to the gas. Thanks to their inverse dependence on $\rho_g$, $\tau$ and $\St$
vary strongly spatially and are much smaller in the dense midplane than in the rarefied upper disk atmosphere.

Dust grains with $\St \ll 1$, well coupled to protoplanetary disk gas, drift through said gas towards pressure maxima with a velocity
\citep[for an in depth derivation of dust velocities in the presence of gas motion and pressure gradients see, e.g.~][]{2016MNRAS.456.3079H}
\EQ
\vv \simeq \frac{\tau}{\rho_g} \nabla p, \label{vdrift}
\EN
where $p$ is the gas pressure.
As a result of \Eq{vdrift}, pressure gradients drive dust drift fluxes
\EQ
\FF_{\text{drift}} = \rho_d \vv \simeq \tau \frac{\rho_d}{\rho_g} \nabla p  \label{Fdr}
\EN
where $\rho_d$ is the dust fluid density. That flux acts to set up dust concentration maxima in gas pressure maxima.
Turbulent diffusion however will act to diffuse away those dust concentration maxima, with a dust diffusive flux
\EQ
\FF_{\text{diff}} =- \rho_g \DD \nabla \left(\frac{\rho_d}{\rho_g}\right) \label{Fdiff}
\EN
where $\DD$ is the diffusion coefficient. Note that diffusion mixes the dust-to-gas mass
ratio
\EQ
\epsilon= \frac{\rho_d}{\rho_g} \label{epsilon_def}
\EN
rather than dust density itself.

In a steady state we have $\FF_{\text{drift}} + \FF_{\text{diff}} = 0$, which implies
\EQ
 \tau \epsilon \nabla p = \rho_g \DD \nabla \epsilon. \label{balance}
\EN
We parameterize the dust diffusion coefficient $\DD=\alpha c_s H$ similar to a \citet{1973A&A....24..337S} $\alpha$-disk,
where $c_s$ is the gas adiabatic
sound speed. The local pressure scale-height
\EQ
H = \frac{1}{\sqrt{\gamma}} \frac{c_s}{\Omega} = \sqrt{\frac{\pi}{8}} \frac{\vth}{\Omega}, \label{HH}
\EN
where $\gamma \simeq 1.4$ is the adiabatic index under standard nebular conditions in chondrule forming regions
of $R \sim 1 - 2.5$\,au \citep{2007ApJ...656L..89B}. Using those definitions for $\DD$ and $H$ we can rewrite Equation~(\ref{balance}) as
\EQ
\nabla \ln \epsilon = \frac{1}{\sqrt{\gamma}} \frac{\St}{\alpha} \nabla \ln p = \frac{S}{\sqrt{\gamma}} \nabla \ln p, \label{balance_raw}
\EN
where we have defined $S \equiv \St/\alpha$ following \cite{2012Icar..220..162J}. Strictly speaking, converting between
an $\alpha$-disk's $\alpha$ and our own requires consideration of both the Schmidt number and any anisotropy
in the turbulent transport but
the difference for dust grains of the size we consider is expected to be of order unity \citep{2006MNRAS.370L..71J}.

However, from \Eq{tauE} it is clear that in general $\tau$, and hence $\St$ and $S$, decreases with increasing pressure.
Accordingly, high pressure regions see smaller drift fluxes, but unchanged diffusive fluxes. Thus it is
useful use the identity
\EQ
p = \frac{\rho_g k_B T}{m_g} = \frac{\pi}{8} \rho_g \vth^2
\EN
(Eq.~\ref{vth}) to rewrite \Eq{balance_raw} as
\EQ
\nabla \ln \epsilon = \frac{\pi}{8 \sqrt{\gamma}} \frac{a \rhoS \Omega \vth}{\alpha} \frac{\nabla p}{p^2} \leq
\frac{\pi}{8 \sqrt{\gamma}} \frac{a \rhoS \Omega_M \vthM}{\alpha_m} \frac{\nabla p}{p^2}, \label{balance_approx}
\EN
where $\Omega_M$, $\vthM$ and $\alpha_m$ are the maximal and minimal $\Omega$, $\vth$, and $\alpha$ over the region of interest,
and we intend to integrate from lower pressure (and thus lower $\epsilon$) to higher pressure (and higher $\epsilon$).
Integrating \Eq{balance_approx} from a point $0$ with $\epsilon=\epsilon_0$, $p=p_0$, $\tau=\tau_0$ and $\vth=\vthO$ to
a point $1$ with $\epsilon=\epsilon_1$ and $p=p_0+\delta p$ we arrive at
\EQ
\ln\left(\frac{\epsilon_1}{\epsilon_0}\right) \leq \frac{\tau_0 \Omega_M}{\sqrt{\gamma} \alpha_m} \frac{\vthM}{\vthO} \left(\frac{\delta p/p_0}{1+\delta p/p_0}\right).
\label{balance_expanded}
\EN
Defining $\St_M \equiv \tau_0 \Omega_M$ and $S_M \equiv \St_M \vthM/\alpha_m \vthO$, \Eq{balance_expanded}
simplifies to
\EQ
\ln\left(\frac{\epsilon_1}{\epsilon_0}\right) \leq \frac{S_M}{\sqrt{\gamma}}\left(\frac{\delta p/p_0}{1+\delta p/p_0}\right)< \frac{S_M}{\sqrt{\gamma}}. \label{lneps}
\EN
We can see that one requires
$S_M >1$ to generate significant variations in the dust-to-gas mass ratio $\epsilon$ along a trajectory of interest.

\section{Dust concentration factors}
\label{DCF}

We consider the case of vertically settled dust further concentrated by a small midplane region of high pressure. When deriving
\Eq{lneps} we integrated from low pressure regions to high pressure regions. However, when the high pressure region
contains a large fraction of the overall dust mass, as occurs with settling, \Eq{lneps} leaves us with an unknown $\epsilon_0$.
We will invoke the constraint that the vertically integrated dust density equals the dust surface density to solve for $\epsilon_0$. In practice, this 
effect mean that even when \Eq{lneps} predicts huge $\epsilon_1/\epsilon_0$ ratios, those ratios say more about the extreme dust depletion in the low pressure
regions than about the dust concentration in the high pressure ones, which remains modest.

\subsection{Settling}

In the case of settling we want to integrate \Eq{balance_raw} from the midplane (i.e. from high pressure to low pressure), and write the mid-plane
dust density in terms of the dust surface density.
\cite{2002ApJ...581.1344T} did this, deriving the equation for the dust density as a function of altitude in a constant $\alpha$, vertically isothermal disk
in vertical hydrostatic equilibrium:
\begin{align}
\rho_g(z) &= \rho_g(0) \exp \left(-\frac{z^2}{2H^2}\right), \\
\rho_d(z) &= \rho_d(0) \exp\left[-\frac{z^2}{2H^2} - \frac{\bS}{\sqrt{\gamma}}\left(\exp\frac{z^2}{2H^2}-1\right)\right],
\end{align}
where $\bS$ is $S$ evaluated at the midplane. Defining
\EQ
I(\bS) \equiv \int_{-\infty}^{+\infty} dz \exp\left[-\frac{z^2}{2H^2} -\frac{\bS}{\sqrt{\gamma}} \left(\exp\frac{z^2}{2H^2}-1\right)\right]
\EN
we get the gas and dust surface densities:
\begin{align}
\Sigma_g &= \sqrt{2 \pi} H \rho_g(0), \\
\Sigma_d &= I(\bS) \rho_d(0).
\end{align}
In the absence of settling (effectively $S=0$), we would have $\epsilon= \bar{\epsilon} \equiv \Sigma_d/\Sigma_g$ as the total system
dust-to-gas mass ratio.

\begin{figure}\begin{center}
\includegraphics[width=0.8\columnwidth]{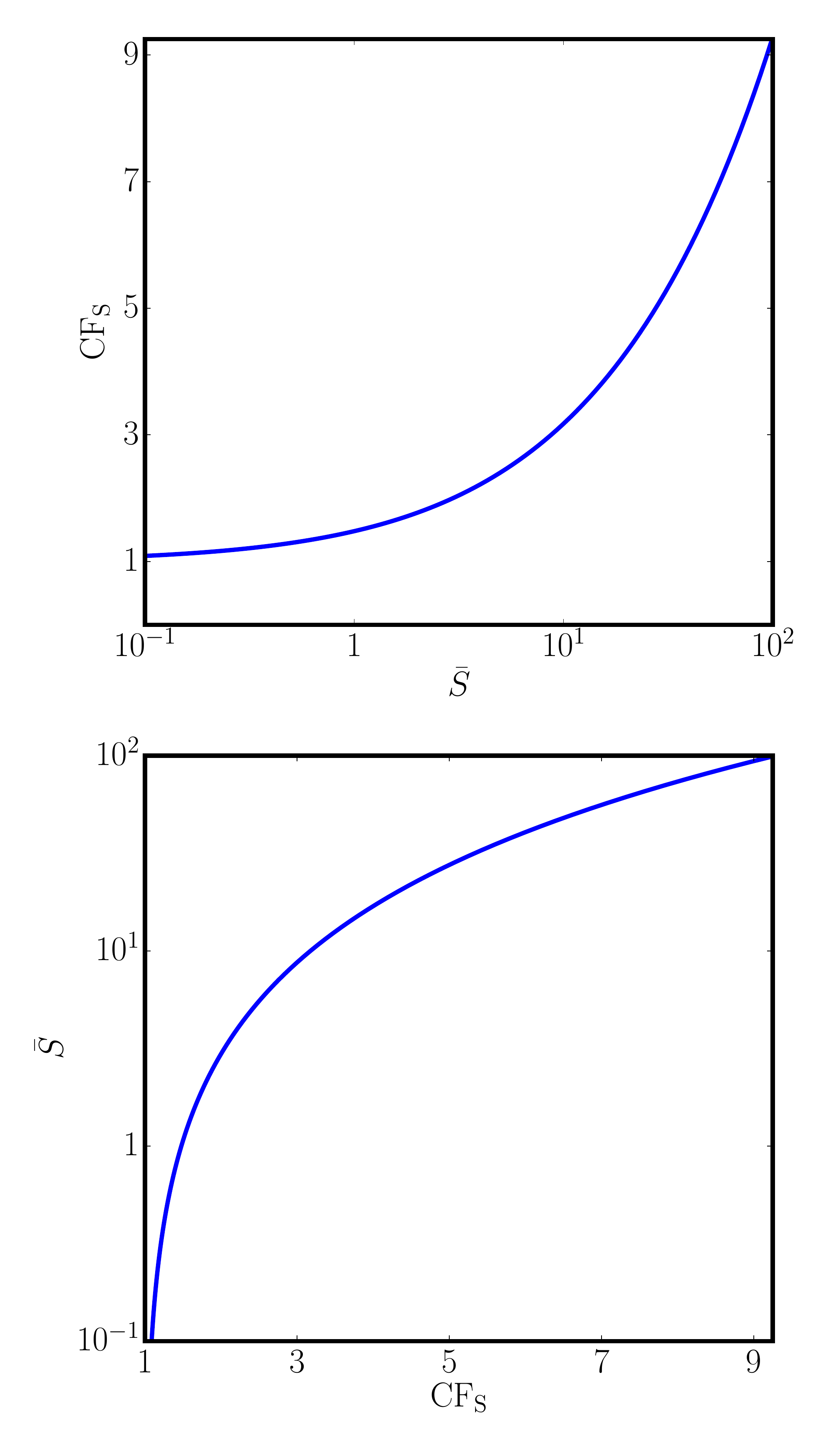}
\end{center}\caption{
Settling concentration factor $\CFS$ as a function of the midplane $\bS$ and vice-versa for $\gamma=1.4$. Settling will only
drive large concentration factors $\CFS \gtrsim10$ for $\bS > 10^2$.
\label{CFS} }
\end{figure}

We can now find that at the midplane 
\EQ
\epsilon(\bS,0) = \frac{\rho_d(0)}{\rho_g(0)} = \frac{\sqrt{2\pi}H}{I(\bS)} \bar{\epsilon}. \label{eps0}
\EN
Using \Eq{eps0} we can usefully define the settling concentration factor at the midplane
\EQ
\CFS(\bS) = \frac{\epsilon(\bS,0)}{\bar{\epsilon}} = \frac{\sqrt{2\pi}H}{I(\bS)} .
\EN
We plot $\CFS$ as a function of $\bS$ and vice-versa in \Fig{CFS}. Note that while settling can lead to large
midplane dust concentrations $\CFS \gtrsim 10$, it can only do so for $\bS \gtrsim 10^2$. Even $\bS=10$ only
leads to $\CFS \sim 3$.

\subsection{Pressure bump}

A radial annulus of high pressure such as a zonal flow \citep{2009ApJ...697.1269J}, or a high pressure anti-cyclonic vortex 
\citep{1995A&A...295L...1B} will further concentrate the already settled particles.
As long as the region is small enough,
the dust mass trapped in the high pressure region will be negligible compared to the total dust mass,
and we can use \Eq{lneps} with $\epsilon_0=\epsilon(\bS,0)$. Thus, we can define the concentration factor
of the pressure bump:
\EQ
\CFP \equiv \frac{\epsilon_1}{\epsilon(\bS,0)} = \exp \left(\frac{\delta p/p_0}{1+\delta p/p_0} \frac{\bS}{\sqrt{\gamma}}\right),
\EN
and the total concentration factor
\EQ
\CFT = \frac{\epsilon_1}{\bar{\epsilon}} = \CFP \times \CFS. \label{CFT_def}
\EN
In \Eq{CFT_def} $\CFT$ is explicitly the dust-to-gas mass ratio in the peak of the pressure bump, at the midplane, normalized to
the overall system dust-to-gas mass ratio.

\begin{figure}\begin{center}
\includegraphics[width=\columnwidth]{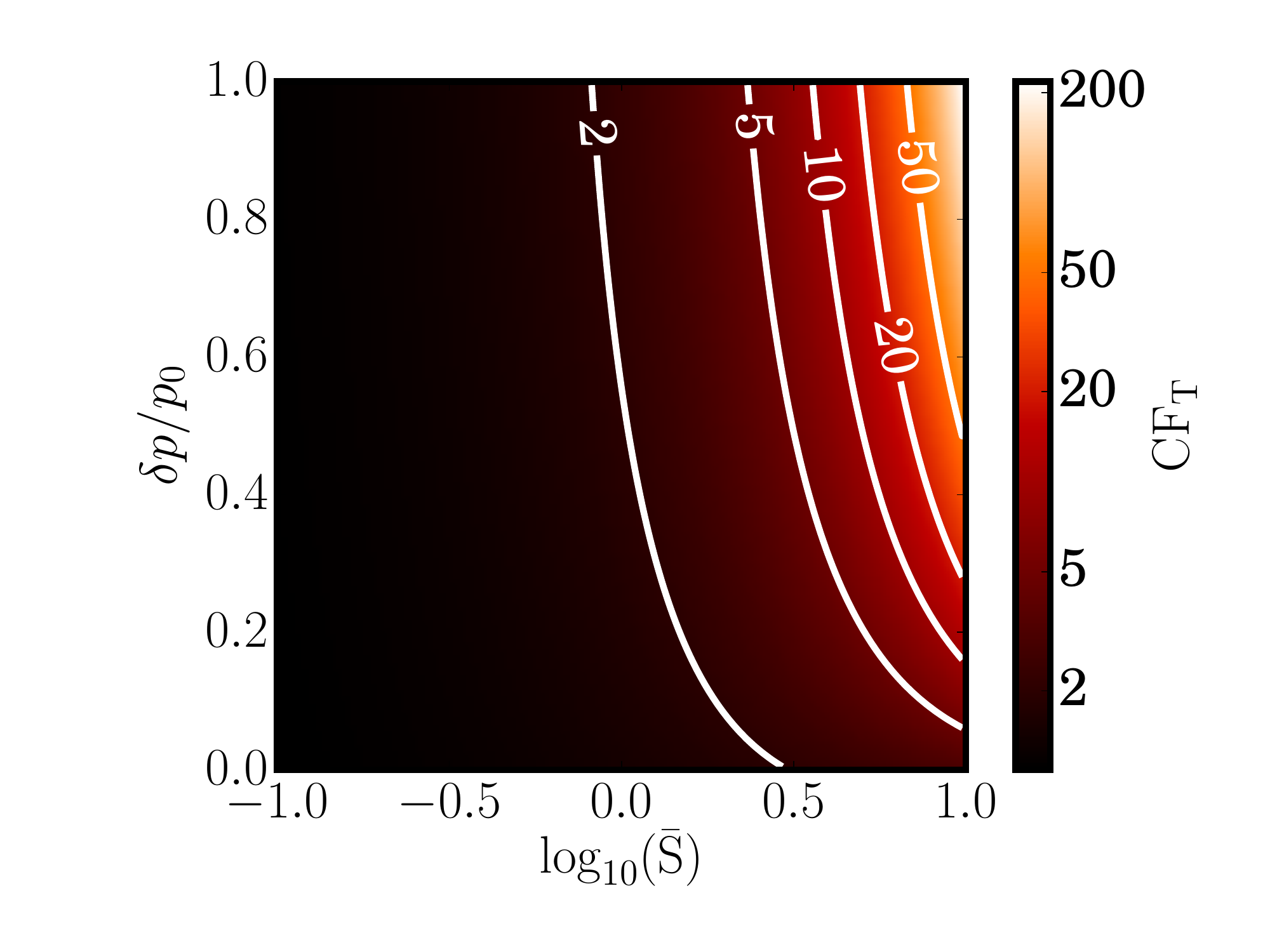}
\end{center}\caption{
$\CFT$ as a function of $\bS$ and $\delta p/p_0$. We can achieve $\CFT=10$ for $\bS=10$ and $\delta p/p_0=0.15$.
\label{CFT} }
\end{figure}

We show $\CFT$ as a function of $\bS$ and $\delta p/p_0$ in \Fig{CFT}, noting that 
$\CFT \simeq 10$ for $\bS=10$ and $\delta p/p_0=0.15$.
The addition of a pressure bump greatly increases the dust concentration
above settling, even though the pressure ratio of the pressure bump is much smaller than that between the disk
midplane and the disk upper atmosphere. This is a consequence of our assumption that the pressure bump is sufficiently spatially constrained
that it contains only a small fraction of the overall dust mass.
This condition can be written as
\EQ
\CFP \times \delta R = \frac{\CFT}{\CFS} \delta R \ll R. \label{radial_extent}
\EN
for a pressure perturbation with radial extent $\delta R$

\section{Application to chondrule precursors}

Our assumption that the pressure bumps are sufficiently spatially localized that they trap only a small fraction of the overall dust mass is highly problematic
for chondrule formation purposes. For such pressure perturbations to play a significant role in chondrule formation, they must have been very tightly correlated with chondrule
formation mechanisms. Nonetheless, we can examine the consequences of \Eq{CFT_def} for chondrules and their precursors.

\subsection{Stokes numbers}

In a vertically isothermal protoplanetary disk, the midplane Stokes number of a dust grain reduces to
\EQ
\St = \frac{\pi}{2} \frac{a \rhoS}{\Sigma_g}. \label{StSig}
\EN
Chondrule sizes vary, with typical radii $a=450,250,135,75\,\mu$m for CV, LL, CM and CO chondrites respectively \citep{Friedrich2014}.
A Hayashi MMSN has surface densities of $\Sigma_g = 1700, 430$\,g\,cm$^{-2}$ at R$=1,2.5$\,au respectively, while chondrules
have solid densities $\rhoS \simeq 3$\,g\,cm$^{-3}$. Those numbers
result in the Stokes numbers given in \Tab{Table_St}. However, chondrule precursors were presumably porous. At constant mass,
$\St \propto \phi^{2/3}$ where $\phi$ is a porous grain's volume filling fraction. We will assume that chondrule precursors were only
modestly collisionally compacted, with $\phi=0.1$ \citep{2010A&A...513A..56G}, resulting in the precursor Stokes numbers also given
in \Tab{Table_St}. All of those Stokes numbers are conspicuously low.

In addition to the condensable enrichment, \cite{2000GeCoA..64..339E} found a need for high pressure gas in chondrule forming regions, with $p_c \sim 10^{-3}$\,atm, with potential trade-offs between gas pressure and condensable enrichments. Using
a mean molecular mass of $2.3$\,amu (i.e.~neglecting any disassociation or ionization)
and a temperature of $1750$\,K, a pressure of $p_c = 10^{-3}$\,atm corresponds to a gas density of $\rho_g=1.6 \times 10^{-8}$\,g\,cm$^{-3}$.
In the limit of the high Mach number required to achieve chondrule forming temperatures from a background of $\sim 300$\,K, and
assuming $\gamma \sim 1.4$,
the pre-shock gas would have had a density of $\rho_g \simeq 3 \times 10^{-9}$\,g\,cm$^{-3}$ and a thermal
speed of $\vth \simeq 1.7 \times 10^5$\,cm\,s$^{-1}$. 
The corresponding Stokes numbers are given in \Tab{Table_St2}, where the R dependence comes
from normalizing the stopping time $\tau$ to the orbital frequency $\Omega(\text{R})$.
Note that if the heating was not due to a shock, the gas would need even
higher densities corresponding to even lower $\St$ values, and that the Stokes numbers are significantly smaller than in the MMSN case.

\begin{table}
\caption{Stokes numbers: MMSN}
\centerline{\begin{tabular}{lcccc}
\hline
Chondrules &  CV & LL & CM & CO \\
\hline
$1$\,au & $1.2\text{e}-4$ & $7\text{e}-5$ & $3.7\text{e}-5$ &  $2.1\text{e}-5$ \\
$2.5$\,au & $4.9\text{e}-4$ & $2.7\text{e}-4$ & $1.5 \text{e}-4$ &  $8 \text{e}-5$ \\
\hline
Precursors &   &  &  &  \\
\hline
$1$\,au & $2.7\text{e}-5$ & $1.5\text{e}-5$ & $8.1 \text{e}-6$ &  $4.5 \text{e}-6$ \\
$2.5$\,au & $1.1 \text{e}-4$ & $5.9\text{e}-5$ & $3.2 \text{e}-5$ &  $1.8 \text{e}-5$ \\
\hline
\end{tabular}} \label{Table_St}
\end{table}

\begin{table}
\caption{Pre-shock Stokes numbers assuming post-shock $p_c=10^{-3}$\,atm}
\centerline{\begin{tabular}{lcccc}
\hline
Chondrules &  CV & LL & CM & CO \\
\hline
$1$\,au & $5.4 \text{e}-5$ & $3.0\text{e}-5$ & $1.6 \text{e}-5$ &  $9 \text{e}-6$ \\
$2.5$\,au & $1.4\text{e}-5$ & $7.6\text{e}-6$ & $4.1 \text{e}-6$ &  $2.3 \text{e}-6$ \\
Precursors &   &  &  &  \\
\hline
$1$\,au & $1.2\text{e}-5$ & $6.5\text{e}-6$ & $3.5 \text{e}-6$ &  $1.9 \text{e}-6$ \\
$2.5$\,au & $3 \text{e}-6$ & $1.6\text{e}-6$ & $8.9 \text{e}-7$ &  $4.9 \text{e}-7$ \\
\hline
\end{tabular}} \label{Table_St2}
\end{table}

\subsection{Midplane $\alpha$ and $\delta p/p_0$}

Even in the case of magnetically dead midplanes \citep{1996ApJ...457..355G} surface layer turbulence will
penetrate to the midplane, driving motions \citep{2009ApJ...704.1239O}. While estimates for the strength
of that turbulence vary, they generally cluster around $\alpha \gtrsim 10^{-4}$
\citep{2009ApJ...704.1239O,2016ApJ...821...80B,2016ApJ...826...18G}. We adopt a conservative
estimate of $\alpha \gtrsim 5 \times 10^{-5}$. Note that even in the absence of other forms of turbulence,
once low-\St\, dust settles to a concentration factor of $\sim 100$, they begin
to drive a sufficiently strong Kelvin-Helmholz instability to prevent further concentration \citep[see e.g.][]{2010M&PS...45..276W}. Thus, even if there is no
external turbulence whatsoever, settling is insufficient to drive high $\CFS$.

Zonal flows generate pressure perturbations on the order of $\delta p/p_0  \lesssim 0.3$ \citep{2009ApJ...697.1269J,2013ApJ...763..117D}.
Rossby waves can generate long-lived vortices at the edges of dead-zones with $\delta p/p_0  \lesssim 1$ \citep{2012ApJ...756...62L}, although the 
perturbations may not stay at their maximum values for long. In both cases however the pressure perturbation has a sufficient radial extent (a few percent
to a few tens of percent the local orbit) that dust concentrations of more than a factor of $10$ will begin to involve
large fractions of the total disk dust mass.

Indeed, that is the general result. Noting that for $\bS \leq 3$ we have $\CFS \leq 2$, and a total
concentration factor of $\CFT > 100$ implies (Equation~\ref{radial_extent}):
\EQ
\frac{\delta R}{R}  \ll \frac{1}{\CFP} = \frac{\CFS}{\CFT} < \frac{1}{50}.
\label{droverR}
\EN
A MMSN has $H/R \sim 1/30$, so to achieve a large $\CFT$ under MMSN-like conditions for $\bS \leq 3$ we need, simultaneously,
$\delta p/p_0 >1$ and $\delta R \ll H$. Such a ``perturbation'' would generate strong enough
outwards pressure forces that the disk's angular momentum would locally cease to increase with radius, violating
the requirement for Rayleigh stability.

\subsection{Implications}

Porous CV chondrule precursors, in extremely quiescent MMSN midplanes ($\alpha = 5 \times 10^{-5}$) could manage $\bS=2$ at $R=2.5$\,au.
Non-porous CM chondrules, in said extremely quiescent MMSN midplanes could approach $\bS=3$ at $R=2.5$\,au. CO chondrules,
much less CO chondrule precursors, struggle to match $\bS=1$. If we want the pressure in chondrule forming regions to approach
$p_c= 10^{-3}$\,atm, then $\bS>1$ is ruled out in all but the most optimistic of cases (extremely quiescent midplanes, close to R$=1$\,au,
non-porous precursors of the largest chondrules).

We can trade off the MMSN approximation for an $\alpha$-disk one by returning to the mass accretion rate for a constant $\alpha$ disk \citep{2002apa..book.....F}:
\EQ
\dot{m} = 3 \pi \alpha_{\text{eff}} c_s \Sigma_g H. \label{dotm}
\EN
Inserting \Eq{dotm} into \Eq{StSig} we arrive at
\begin{align}
S \simeq 0.76 &\left(\frac{a \rhoS}{0.135 \text{ g cm}^{-2}}\right) \left(\frac{\dot{m}}{10^{-8} \text{M}_\odot\text{ yr}^{-1}}\right)^{-1} \nonumber \\
&\times \left(\frac{\alpha}{\alpha_{\text{eff}}}\right)^{-1} \left(\frac{R}{2.5\text{ au}}\right)^{3/2},
\end{align}
where we have normalized $a \rhoS$ to large non-porous CV chondrules. 
Quiescent T Tauri stars accrete at around $10^{-8}$ M$_\odot$ yr$^{-1}$
\citep{2013ApJ...767..112I} although values closer to $10^{-9}$ M$_\odot$ yr$^{-1}$ occur. While the link between the midplane
dust diffusion coefficient $\alpha$ and the accretion driving $\alpha_{\text{eff}}$ is unclear, it is difficult to imagine $\alpha < 0.1 \alpha_{\text{eff}}$.
Thus, we are still left with CO chondrule precursors having $\bS < 3$ even in the most optimistic of cases.

\begin{figure}\begin{center}
\includegraphics[width=\columnwidth]{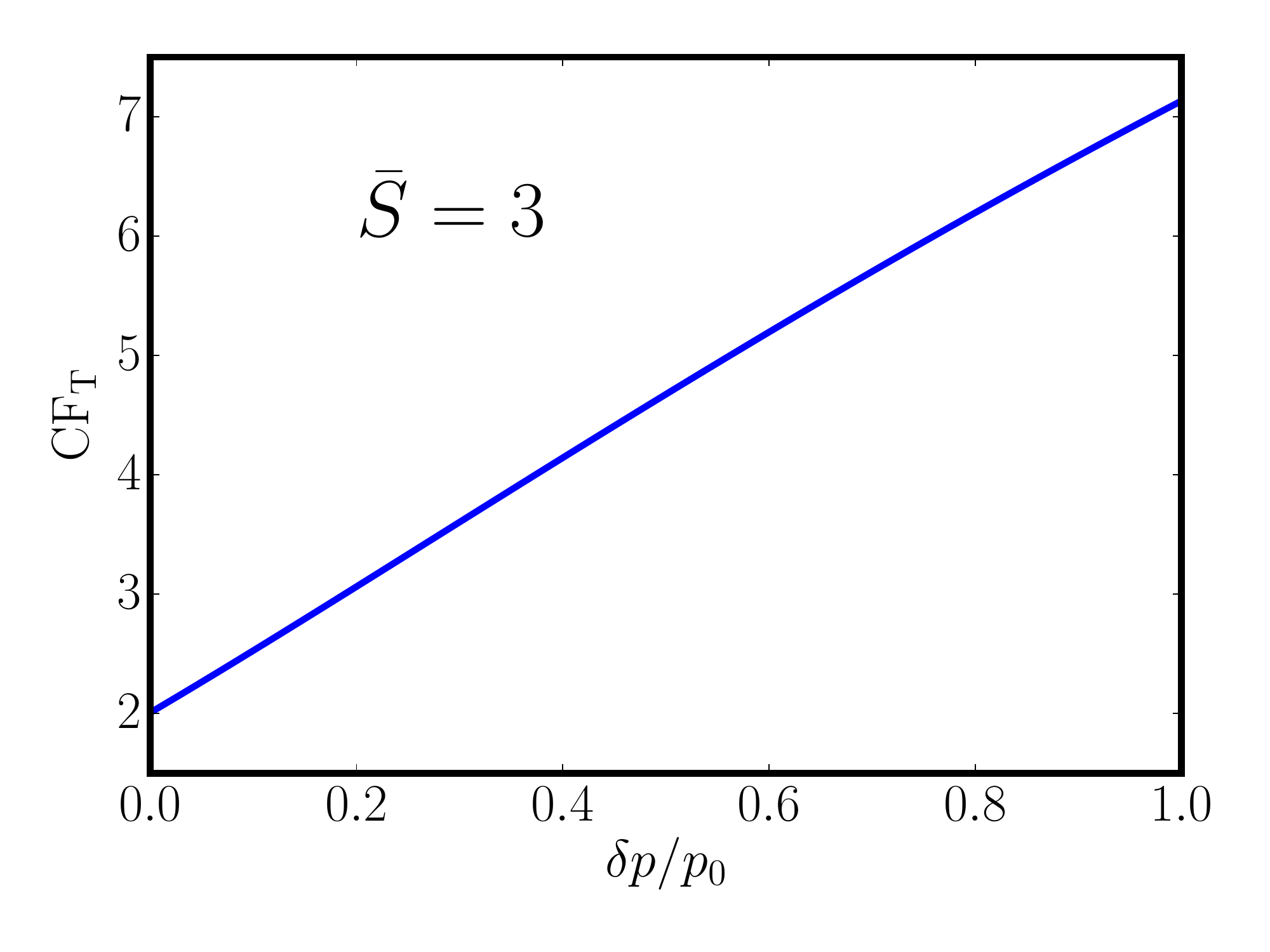}
\end{center}\caption{
$\CFT$ as a function of $\delta p/p_0$ for $\bS=3$.
\label{S3} }
\end{figure}

We cannot achieve even merely $\CFT=10$ with these maximal estimates for $\bS \leq 3$ as shown in
\Fig{S3}, noting further that $\delta p/p_0$ is not expected to be large. Thus, we find ourselves with
predictions of $\CFT < 10$ for chondrules and chondrule precursors under all conditions except perhaps for CV chondrules (not precursors) at $R=2.5$\,au
in an extremely quiescent disk ($\alpha=5 \times 10^{-5}$) with very strong long-lived pressure perturbations ($\delta p/p_0=0.15$).
Some further intermittent concentration is possible: perhaps a factor of $6$ from a strong shock, and a factor of $2$ from preferential concentration
 \citep{2013MNRAS.432.1274H}, but final factors of hundreds, or even tens, are not expected.

\section{Further complications}

\subsection{Non-concentrated dust}

In the calculations above we explicitly assumed that most of the dust mass is not in the regions of peak concentration. In that model,
most heating events strong enough to melt chondrules should have hit non-concentrated dust
unless the heating mechanism was tightly linked to the pressure perturbation concentrating the dust.
Requiring that large fractions of the dust mass
be in dense clumps greatly increases the difficulty in concentrating the dust, and is inconsistent with observations: we see dust almost
everywhere in protoplanetary disks with gaps being the exciting exception \citep{2015ApJ...808L...3A}.
Relaxing this assumption reduces the amount of dust in the regions of peak concentration: there is not an arbitrary amount of
dust available to parcel out to the pressure maxima. This is the reason why the vast pressure ratios between the upper disk atmosphere and the disk
midplane only generate dust concentration factors comparable to much more moderate midplane pressure perturbations of tens of percents.

\subsection{Concentrated dust}

While as we have shown one does not expect strong concentrations of chondrules or their precursors we can also examine the conditions
required for strong concentrations. From \Fig{CFT}, we could reasonably
get $\CFT \gtrsim 10$ for $\bS \gtrsim 10$ and $\delta p/p_0=0.15$. Given a moderate estimate for a laminar disk midplane of $\alpha=10^{-4}$ that implies $\St \gtrsim 10^{-3}$.
\cite{2016arXiv161107014Y} find that those concentrations and Stokes numbers lead to the streaming instability \citep{2005ApJ...620..459Y}, which can
rapidly create dense, gravitationally
unstable dust clouds. Thus, concentrations $\CFT \gtrsim 10$ are not expected to be long lived, but instead
to be a brief stage in the process of planetesimal formation. In that case, it becomes difficult to imagine why chondrule forming events would strike only
during such a brief window in the dust's life, and it is difficult to see how matrix material (much of which stayed well below chondrule forming temperatures) would have 
had time to be evenly mixed in with
the newly formed chondrules before planetesimal formation finalized.

Strong radial pressure perturbations concentrate
dust grains in pressure maxima, regions with zero radial pressure gradients, and so do not host the streaming instability. However,
this also means that concentrated dust layers do not trigger Kelvin-Helmholtz instabilities, allowing the classical Safronov-Goldreich-Ward
gravitational instability \citep{1969edo..book.....S,1973ApJ...183.1051G}. Thus, high dust concentration requirements
find themselves on the horns of a dilemma: chondrule precursors are not expected to have been strongly concentrated, but even if they were, those
concentrations are not expected to have lasted long enough to
allow chondrule formation events to preferentially strike them. 

\subsection{Concentration times scales}

Our concentration analyses in Sections~\ref{MDC} and \ref{DCF} assume equilibrium between dust drift and turbulence. The
vertical settling time of $t \sim 1/ 2 \pi \St$ local orbits is short compared to disk lifetimes for all but the smallest grains
considered in Table~\ref{Table_St} or \ref{Table_St2}, but the same is not true for concentration in radial pressure bumps.
Radial drift in an MMSN-like disk occurs at a speed of approximated
\EQ
v_{\text{drift}} \simeq 50 \St \text{ m/s} \simeq 10^{-2} \St \text{ au/yr}.
\EN
Assuming dust concentrates from an annulus of
width $\triangle R$ final annulus of width $\delta R$, to achieve the concentration factor $\CFP = 50$ assumed in \Eq{droverR}
one requires
\EQ
\frac{\triangle R}{R} \simeq \left[\sqrt{100 \frac{\delta R}{R} + 1} - 1\right]
\label{triangleR}
\EN
where we have already assumed $\delta R \ll R$. For the smallest reasonable annulus width $\delta R /R \simeq H /R  \sim 1/30$,
\Eq{triangleR} implies $\triangle R \simeq R > 1$\,au and hence a drift time greater than $100 \St^{-1}$\,years. For the values in 
Tables~\ref{Table_St} and \ref{Table_St2}, this ranges from 200 kyr in the most generous case up to 200 myr in the least
generous case. It is hard to imagine a radial pressure perturbation remaining unmodified for such a long time even in the most
optimistic of cases.

\section{Conclusions}

We have examined the requirement for strong dust concentration in protoplanetary disks finding that chondrule precursors
are not expected to have been concentrated even by a single order of magnitude. That is in strong tension with cosmochemical analyses that
find chondrule
forming events had to have occurred in regions with dust concentrations of factors of hundreds to thousands \citep{2000GeCoA..64..339E,2015GeCoA.148..228T}.
While the details depend on the overall disk density, the higher the density, the harder it is to concentrate dust.
 We have examined both the $p = 10^{-3}$\,atm case preferred by cosmochemists and the significantly lower density
 Hayashi MMSN Solar Nebula \citep{2000GeCoA..64..339E, 1981PThPS..70...35H}, finding that both are too dense to allow significant dust concentration. While
sufficiently rarified gas to allow high dust concentration factors are expected late in a protoplanetary disk's life,
they are faced with a separate problem. Dynamical theory predicts that if and when extremely high dust concentrations are achieved, they are but
brief stages in dust evolution, lasting significantly less than $1000$\,yr, and culminating
in the formation of planetesimals through collective instabilities such as the streaming instability \citep{2005ApJ...620..459Y}.
Even if such concentrations of chondrule precursors were
possible it is therefore difficult to imagine how they could have been sufficiently correlated with chondrule forming events.

Most of the disk cannot (by definition) have extremely high dust concentrations. High dust concentrations in chondrule forming events therefore requires
that chondrule forming events either have caused or been caused by high dust concentrations, and that those concentrations not have led
to rapid planetesimal formation or accretion onto planetesimals. Dynamical concentration of chondrule sized dust appears
ruled out outside of the very end of the disk's lifetime, and when it is possible, dust concentration naturally leads to planetesimal formation.
The goal posts for nebular high dust concentration chondrule formation are thus extremely narrow. Chondrite classes are sufficiently
numerous that absent a large chondrite class-to-planetesimal ratio \citep[e.g.][]{2013AREPS..41..529W}, invoking
extremely rare conditions is disfavored. We seem to be left with discarding either ``nebular'' or ``high concentration'' in
chondrule formation.

This concentration contradiction provides an exciting way to link cosmochemistry to disk dynamics and to learn from both.
The cosmochemical requirements are sufficiently distant from the dynamical predictions, and turbulent diffusion of passive scalars is a sufficiently well
understood problem, that the cosmochemical constraints need to be revisited, perhaps varying dust compositions \citep{2016GeCoA.172..322E}
or appealing to kinetic effects \citep{1996GeCoA..60.1445N}. If no alternatives with strongly lowered condensable concentrations can be found,
we will need to rethink our understanding of protoplanetary disks and dust transport therein at a fundamental level, or to adopt non-nebular
models of chondrule formation \citep{1953GeCoA...4...36U}.

\section*{Acknowledgements}
We thank Stu Weidenschilling and an anonymous referee for their useful comments, suggestions, and elaborations on the difficulties we present here.
The research leading to these results was funded by NASA OSS grant NNX14AJ56G (AH) and EW grant NNXI6AD37G (DE).

\bibliography{Conc}

\end{document}